\documentclass[nofootinbib,preprint,aps,showpacs]{revtex4}

\usepackage{graphicx,epsfig}

\begin{document}

\title{Modification of jet-like correlations in 
Pb-Au collisions at 158$A$~GeV/$c$}

\def\rez{$^{(1)}$}
\def\gsi{$^{(2)}$}
\def\fra{$^{(3)}$}
\def\dub{$^{(4)}$}
\def\mpi{$^{(5)}$}
\def\hei{$^{(6)}$}
\def\tud{$^{(7)}$}
\def\emm{$^{(8)}$}
\def\fia{$^{(9)}$}
\def\wei{$^{(10)}$}
\def\mun{$^{(11)}$}
\def\sun{$^{(12)}$}
\def\cer{$^{(13)}$}
\def\bnl{$^{(14)}$}

\author{D.~Adamov\'a\rez, 
G.~Agakichiev\gsi, 
D.~Anto\'nczyk\fra,
H.~Appelsh\"auser\fra, 
V.~Belaga\dub, 
J.~Biel\v{c}\'ikov\'a\mpi\hei,
P.~Braun-Munzinger\gsi\tud\emm\fia,
O.~Busch\hei, 
A.~Cherlin\wei, 
S.~Damjanovi\'c\hei, 
T.~Dietel\mun, 
L.~Dietrich\hei, 
A.~Drees\sun, 
W.~Dubitzky\hei,
S.\,I.~Esumi\hei, 
K.~Filimonov\hei, 
K.~Fomenko\dub,
Z.~Fraenkel\wei$^{\dagger}$, \footnotetext{$^{\dagger}$deceased} 
C.~Garabatos\gsi, 
P.~Gl\"assel\hei, 
J.~Holeczek\gsi, 
M.~Kalisky\mun,
S.~Kniege\fra,
V.~Kushpil\rez, 
A.~Maas\gsi, 
A.~Mar\'{\i}n\gsi, 
J.~Milo\v{s}evi\'c\hei,
A.~Milov\wei, 
D.~Mi\'skowiec\gsi, 
Yu.~Panebrattsev\dub, 
O.~Petchenova\dub, 
V.~Petr\'a\v{c}ek\hei, 
A.~Pfeiffer\cer,
M.~P\l{}osko\'n\fra, 
J.~Rak\mpi, 
I.~Ravinovich\wei, 
P.~Rehak\bnl,
H.~Sako\gsi, 
W.~Schmitz\hei, 
S.~Schuchmann\fra,
S.~Sedykh\gsi, 
S.~Shimansky\dub, 
J.~Stachel\hei, 
M.~\v{S}umbera\rez, 
H.~Tilsner\hei, 
I.~Tserruya\wei, 
J.\,P.~Wessels\mun, 
T.~Wienold\hei, 
J.\,P.~Wurm\mpi, 
W.~Xie\wei, 
S.~Yurevich\hei, 
V.~Yurevich\dub \\
(CERES Collaboration)}

%\address{
\affiliation{
%\rez NPI ASCR, \v{R}e\v{z}, Czech Republic\\
\rez Nuclear Physics Institute ASCR, 25068 \v{R}e\v{z}, Czech Republic\\
%\gsi GSI Darmstadt, Germany\\
\gsi GSI~Helmholtzzentrum~f\"{u}r~Schwerionenforschung~GmbH,~D-64291~Darmstadt,~Germany\\
%\fra Frankfurt University, Germany\\
\fra Institut f\"{u}r Kernphysik, Goethe Universit\"{a}t Frankfurt,~D-60486 Frankfurt,~Germany\\
%\dub JINR Dubna, Russia\\
\dub Joint Institute for Nuclear Research, 141980 Dubna, Russia\\
%\mpi MPI, Heidelberg, Germany\\
\mpi Max-Planck-Institut f\"{u}r Kernphysik, D-69117 Heidelberg, Germany\\
%\tud Technische Universit\"{a}t Darmstadt, Germany\\
\tud Institut f\"{u}r Kernphysik, Technische Universit\"{a}t Darmstadt,
D-64289 Darmstadt, Germany\\
%\emm EMMI
\emm ExtreMe Matter Institute EMMI,
GSI~Helmholtzzentrum~f\"{u}r~Schwerionenforschung~GmbH,~D-64291~Darmstadt,~Germany\\
%\fia FIAS
\fia Frankfurt Institute for Advanced Studies, Goethe Universit\"{a}t Frankfurt,~D-60438 Frankfurt, Germany\\
%\hei Heidelberg University, Germany\\
\hei Physikalisches Institut, Ruprecht-Karls Universit\"{a}t Heidelberg, D-69120
Heidelberg, Germany\\
%\wei Weizmann Institute, Rehovot, Israel\\
\wei Weizmann Institute, Rehovot 76100, Israel\\
%\mun M\"unster University, Germany\\
\mun Institut f\"{u}r Kernphysik, Westf\"{a}lische Wilhelms-Universit\"{a}t M\"unster, D-48149 M\"unster,~Germany\\%\sun SUNY at Stony Brook, NY, U.S.A.\\
\sun Department of Physics and Astronomy, State University of
New York--Stony Brook, Stony Brook, New York 11794-3800\\
%\cer CERN, Geneva, Switzerland\\
\cer CERN, 1211 Geneva 23, Switzerland\\
\bnl Brookhaven National Laboratory, Upton, New York 11973-5000\\
}

\begin{abstract}
  Results of a two-particle correlation analysis of high-$p_t$ charged particles
  in Pb-Au collisions at 158$A$ GeV/$c$ are presented. The data have been recorded
  by the CERES experiment at the CERN-SPS. The correlations are
  studied as function of transverse momentum, particle charge and collision centrality.
  We observe a jet-like structure in the vicinity of a high-$p_t$ trigger particle 
  and a broad back-to-back distribution. The yields of associated particles per
  trigger show a strong dependence on the trigger/associate charge combination.
  A comparison to PYTHIA confirms the jet-like pattern at the near-side but suggests
  a strong modification at the away-side, implying significant energy transfer of
  the hard-scattered parton to the medium.
\end{abstract}

\maketitle 

\section{Introduction}

The study of high-energy particle jets in ultrarelativistic heavy-ion 
collisions provides access to the properties of the dense medium formed 
in such reactions. In contrast to the situation in 
elementary collisions, the hard-scattered 
partons have to traverse the surrounding strongly interacting matter  
which leads to significant energy loss and to a modification
of the final state particle jet~\cite{XNW1,XNW2,BDMPS1,BDMPS2}. 
Particularly strong energy loss 
of the parton in the medium is expected if the matter is deconfined
and has free color charge carriers.
This may lead to a significant suppression of high transverse momentum ($p_t$)
hadrons compared to expectations based on perturbative QCD,
if a Quark-Gluon Plasma (QGP) has been formed. 
Furthermore, the energy which has been radiated off the parton 
prior to hadronization may lead to enhanced soft particle production in 
the angular vicinity of the parton. Possible mechanisms like Cerenkov gluon
radiation~\cite{dremin,koch}, shock wave formation~\cite{stoecker,shuryak} 
and large angle gluon radiation~\cite{vitev} have been 
discussed.

A strong suppression of high-$p_t$ hadrons in Au-Au collisions at 
$\sqrt{s_{\rm NN}}=200$~GeV compared to p-p has been
reported~\cite{brahms-wp,phobos-wp,star-wp,phenix-wp,phe1,phe2,star1,phe3}. 
Similarly, early studies of azimuthal correlations revealed 
a significant reduction of the di-jet over jet rate in such collisions
which became manifest in a dramatic disappearance of the back-to-back 
correlation to a high-$p_t$ trigger particle~\cite{star2}.
These observations have been interpreted as a strong indication for 
the formation of a dense colored 
medium in nuclear collisions at the Relativistic Heavy Ion Collider (RHIC).

Pioneering studies of two-pion correlations at the Super Proton Synchrotron
(SPS) have indicated
a significant broadening of the back-to-back correlation in Pb-Au at 
$\sqrt{s_{\rm NN}}=17.3$~GeV
but no disappearance~\cite{ceres1}. 
A very similar broadening has 
also been observed at RHIC~\cite{star3,phe4,phe5} 
when the analysis was extended to lower transverse momentum, 
where the back-to-back correlation 
exhibits a double-humped structure with a local minimum around
$\Delta \phi = \pi$. These findings have been connected to strong
final state interactions of the hard-scattered parton with the matter,
indicating the occurence of significant energy transfer from the
parton to the medium.
Final conclusions on the relevant mechanisms at work, however, have not
yet been drawn, and more conventional scenarios like jet-redirection
through multiple scattering have not yet been ruled out. 
In this Letter, we present results of a two-particle correlation analysis
in Pb-Au collisions at 158~$A$~GeV/$c$.
These data represent a high-statistics measurement of jet-like particle
correlations at $\sqrt{s_{\rm NN}}=17.3$~GeV which, in addition to existing
results at higher collision energy, may help to disentangle the underlying
mechanisms of jet modifications by means of a systematic energy scan.

\section{Experiment and Data Selection}

The data were recorded with the CERES experiment at the CERN-SPS. In 1998, the 
spectrometer was upgraded by a cylindrical Time Projection Chamber 
(TPC) with radial drift field, matching the spectrometer acceptance in
pseudorapidity $2.1 < \eta < 2.6$, and with
full coverage in azimuth~\cite{ceres2}. 
The TPC is hence
ideally suited for studies of azimuthal correlations in the mid-rapidity region
of heavy-ion collisions at 158$A$~GeV/$c$ ($y_{\rm mid}=2.91$). Charged 
particle momenta can be reconstructed from the track curvature in the magnetic
field. Measurement of the particle trajectories in up to 20 planes along the 
beam direction results in a momentum resolution of 
$\Delta p/p = ((2\%)^2+(1\% \cdot p~)^2)^{1/2}$, with $p$ in GeV/$c$. 

The present analysis is based on $3\cdot10^7$ Pb-Au collisions recorded 
with the TPC in the year 2000. The event sample is analyzed in three bins of 
centrality (0-5\%, 5-10\%, and
10-20\% of the most central of the total geometric cross section $\sigma_{\rm geom}$,
respectively). The centrality is determined via the multiplicity of charged
particles measured in the TPC and a telescope of Silicon Drift Detectors 
(SDD) close to the target~\cite{misko-qm}\footnote{For the trigger-induced 
distribution of the events within the centrality bins see~\cite{misko-qm}.}.

The charged particle tracks used for this analysis are selected from the polar angle range
$0.14 < \theta < 0.24$. To provide sufficient momentum resolution, the tracks
are required to have at least 12 out of 
20 possible points used in the momentum fit. 
The accepted tracks are subdivided into a trigger and an associate sample,
based on the transverse momentum $p_t$ of the tracks. In a given event,
the trigger is the 
particle of highest transverse momentum within the trigger $p_t$ range.
Except for the study of the transverse momentum dependence, 
we choose trigger $(T)$ and associate $(A)$ particles in the ranges
$2.5 < p_{t}(T) < 4.0$~GeV/$c$ 
and $1.0 < p_{t}(A) < 2.5$~GeV/$c$, respectively.
Using this selection, the overall number of events, trigger particles and associated
particles per trigger in the different centrality classes are given
in Table~\ref{tab1}.

The single track efficiency for this set of cuts has been 
determined by a Monte Carlo procedure 
where simulated tracks were embedded into real data events. 
This method 
provides a detailed simulation of the TPC response to charged particles 
in a realistic track density environment. 
The efficiency was found to be $\epsilon = 0.78$ 
for particles with $p_t > 1.0$~GeV/$c$,
with no significant dependence on rapidity, transverse momentum and 
charge of the particle.

Within the same Monte Carlo study, the momentum resolution was determined
as function of the polar angle $\theta$, the number of fitted points in
the TPC, the transverse momentum, and the charge of the particle.
For typical trigger and associate ranges ($2.5 < p_{t}(T) < 4.0$~GeV/$c$ 
and $1.0 < p_{t}(A) < 2.5$~GeV/$c$) as used in this analysis,
the mean transverse momentum of trigger and associated particles 
is $\langle p_t(T)\rangle = 2.88$~GeV/$c$ and 
$\langle p_t(A)\rangle = 1.3$~GeV/$c$, respectively.
The typical transverse momentum resolution is
$\Delta p_t/p_t = 7.5\%$ for the trigger particles and $\Delta p_t/p_t = 4.0\%$
for associated particles.

The resolution of the azimuthal and polar angles is better than 3~mrad and 
1~mrad, respectively, for particles with $p_t > 1$~GeV/$c$ and hence negligible
for the shape and magnitude of the angular correlations presented in this
study.

The limited efficiency to reconstruct two tracks with small angular
separation was studied by dividing the distribution of opening angles 
of pairs from the same event by a reference distribution obtained
by combining particles from different events. For opening angles
greater than 15~mrad the two-track inefficiency is less than 1\%. A corresponding
cut is applied to all signal- and mixed-event distributions accumulated
in this analysis.

\begin{center}
\begin{table}
\caption{Number of events $N_{\rm ev}$, triggers $N_T$ and associated particles
per trigger $N_A/N_T$ in the different centrality bins.}
~
\begin{tabular}{c c c c}
\hline \hline
$~~~\sigma/\sigma_{\rm geom}(\%)~~~~ $ & ~~~~$N_{\rm ev}$~~~~~ & ~~~~~~~$N_T$~~~~~~~~~~ & $N_A/N_T$ \\
\hline
0.0 - 5.0    & $1.5\cdot 10^7$ & $2.5\cdot10^6$ & 14.5 \\
5.0 - 10.0   & $9.7\cdot 10^6$ & $1.4\cdot10^6$ & 12.2 \\
10.0 - 20.0  & $3.7\cdot 10^6$ & $4.2\cdot10^5$ & 9.9\\
\hline \hline
\end{tabular}
\label{tab1}

\end{table}
\end{center}

\section{Data Analysis}

Jet-like correlations are studied by measuring, for a trigger particle
and all associated particles in an event, the distribution
$S(\Delta \phi)$ where $\Delta \phi = \phi_1 - \phi_2$ is the 
difference in the azimuthal angle between trigger 
and associated particle. 
Correlations arising due to slight non-uniformities of the single-track acceptance
are accounted for by division by a mixed-event distribution $B(\Delta \phi)$,
where trigger and associated particles are taken from different events of the same 
centrality class. To reduce the statistical uncertainty in the background
distribution, ten events were mixed for each signal event.

The normalized correlation function
\begin{equation}
	C_2(\Delta \phi) = \frac{\int{B(\Delta \phi')d(\Delta \phi')}}
				{\int{S(\Delta \phi')d(\Delta \phi')}}
	 \cdot \frac{S(\Delta \phi)}{B(\Delta \phi)}
\end{equation}
contains physical correlations from jets as well as azimuthal anisotropies
due to the finite impact parameter of the collision. To separate the 
jet-like correlations from the bulk anisotropies known as elliptic flow,
we assume that each particle pair can be attributed to either of the two
contributions. In this two-source approach the correlation function 
is decomposed into the jet-like contribution $C_{2,jet}$ and a contribution
arising due to elliptic flow,
\begin{equation}
	C_2(\Delta \phi) = C_{2,jet}(\Delta \phi)+
		b\cdot(1+2\langle v_2^T \rangle 
		\langle v_2^A\rangle \cos(2\Delta \phi)), 
\end{equation}
where $\langle v_2^T \rangle$ and $\langle v_2^A\rangle$ are the average 
elliptic flow coefficients determined in the trigger and associate 
$p_t$ range, respectively. Assuming that the jet yield vanishes at its
minimum (the ZYAM assumption~\cite{phe5,ZYAM}),
the elliptic flow contribution is adjusted by variation of $b$ to a 
polynomial fit of $C_2(\Delta \phi)$ and subtracted. The polynomial 
fit of $C_2(\Delta \phi)$ is used in order to be less sensitive to point-by-point 
fluctuations of the correlation function.
We obtain the conditional yield $ \hat{J}_2(\Delta \phi) $ 
as the number of jet-associated particles
per trigger:
\begin{equation}
  \hat{J}_2(\Delta \phi) \equiv \frac{1}{N_T}
	\frac{dN_{TA}}{d\Delta\phi}=  \frac{1}{\epsilon} 
  \frac{C_{2,jet}(\Delta\phi)}{\int{C_2(\Delta \phi')d(\Delta \phi')}}
  \frac{N_A}{N_T},
\end{equation}
where $N_T$ and $N_A$ are the total numbers of triggers and associates,
and $N_{TA}$ is the number of jet-associated particles with the trigger 
after subtraction of the flow-modulated background. The yield is corrected
for the single track efficiency $\epsilon = 0.78$.

The $p_t$-dependent elliptic flow coefficients in the trigger and
associated ranges have been determined by a centrality dependent
analysis of azimuthal correlations~\cite{milo}. For these studies,
the usual reaction plane method has been employed. 
The analysis has been
performed for all charged particles, 
as well as for positively or negatively charged particles only.
Auto-correlations have been removed by exclusion of particle $i$ from
the reaction plane calculation. In events with jets, however, particles
are clustered and the reaction plane could be biased to the jet axis
even after removal of particle $i$ from the reaction plane calculation.
To study the impact of this potential bias, separate analyses have been 
performed 
for events with and without a trigger particle. In addition, the results
of the triggered events have been compared to an analysis where only
particles with $p_t < 1$~GeV/$c$ have been used for the calculation of
the reaction plane. In all cases, the extracted flow coefficients after
correction for the resolution of the reaction plane
agree within their statistical uncertainties.

\section{Results and discussion}
\begin{figure}[ht]
\begin{center}
\includegraphics[width=13.0cm]{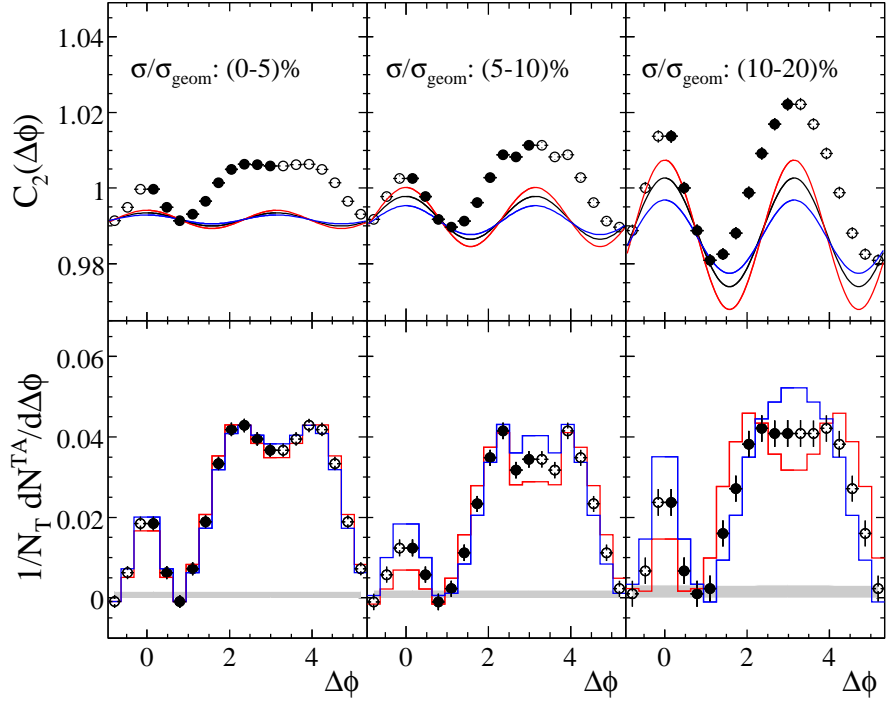}
   \caption{(Color online)
	Two-particle correlation function 
	(upper row) and conditional jet-associated
	yield (lower row) in Pb-Au collisions at 158$A$ GeV/$c$ for
	three ranges in collision centrality. The open symbols are reflected
	about $\Delta\phi = \pi$. The blue
	and red lines and histograms, and the grey band indicate 
	the systematic uncertainty 
	in the flow subtraction and determination of the $b$-parameter, 
	respectively.}
\label{fig1}
\end{center}
\end{figure}

The upper row of Fig.~\ref{fig1} shows the correlation function
$C_2(\Delta \phi)$ and the estimated contribution from elliptic flow
in three different bins of collision centrality. 
The blue and
red lines indicate the uncertainty arising from the systematic error
of the $v_2$ coefficients which are about 10\% and
25\% in the associate and trigger $p_t$ range, respectively. 
These errors have been added linearly 
because a strong correlation is assumed, hence leading to a conservative
estimate of the resulting uncertainty bands. The lower row in Fig.~\ref{fig1}
shows the extracted conditional yields, corrected for the single track
efficiency. At all centralities, we observe a narrow peak at the near
side ($\Delta \phi \approx 0$) and a broad distribution at the away-side
($\Delta \phi \approx \pi$). 
No significant change of yield and shape
on the near- and away-side is observed within the centrality range
under study. The narrow peak on the near-side is indicative of jet-like 
correlations to a high-$p_t$ trigger particle. The away-side is broad,
reminiscent of similar observations in Au-Au collisions at higher collision 
energies, where
the shape is in contrast to observations in p-p collisions at the same
energies. These findings have been discussed in the context of strong
final state interactions of the hard-scattered parton in a dense medium.

%%% figure 1.....

In the most central collisions, we observe that the shape of the distribution 
is significantly non-Gaussian and exhibits a local minimum 
around $\Delta \phi = \pi$. 
The PHENIX experiment at RHIC reported similar 
measurements in Au-Au collisions 
at $\sqrt{s_{\rm NN}}=200$~GeV and $\sqrt{s_{\rm NN}}=62.4$~GeV~\cite{phe6}. 
PHENIX covers approximately the same pseudorapidity 
acceptance and uses the 
same trigger and associate $p_t$ ranges as in the present analysis. 
We observe qualitative agreement with these RHIC results, 
in particular
with respect to the shapes of the near- and away-side distributions. 
Quantitatively, a weaker near-side correlation is observed at SPS, which 
can be explained by a lower average jet energy at a given
trigger threshold. For the same reason, the away-to-near-side yield ratio at SPS is 
larger than at RHIC. 
The latter observation is also in agreement with the 
expectation of a larger 
di-jet to mono-jet rate at SPS in the acceptance, since a given trigger 
threshold biases the measurement towards larger 
$x_t=2 \cdot p_t/\sqrt{s_{\rm NN}}$ at lower beam energy,
and hence to more symmetric parton-parton collisions around mid-rapidity. 

\begin{figure}[ht]
\begin{center}
\includegraphics[width=16.0cm,angle=0]
	{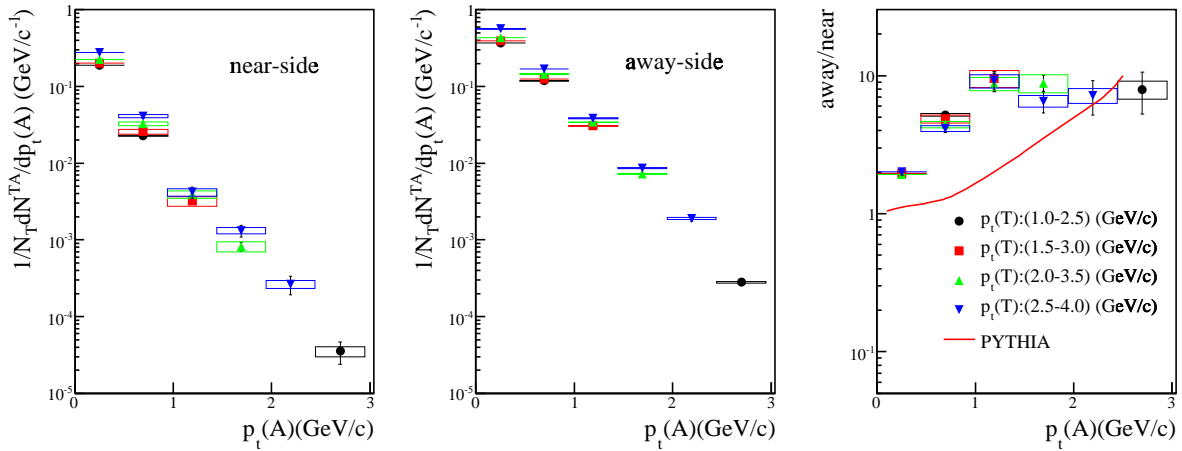}
   \caption{(Color online)
	Jet-associated yield per trigger as function of the associated 
   	$p_t$
	for the near-side (left panel) and the away-side (middle panel)
	in central Pb-Au collisions at 158$A$~GeV/$c$. 
	The right panel shows the ratio of the integrated jet-associated yield distributions 
	away/near in comparison to PYTHIA predictions. The data shown elsewhere in this paper
	are all obtained using $1<p_t(A) <2.5$~GeV/$c$ for the associated particles.}
\label{fig2}
\end{center}
\end{figure}

The yields of particles associated with a high-$p_t$ trigger in central events
have been studied 
in different windows of the trigger and associated $p_t$. For this study,
associated particles at $|\Delta\phi| < 1$ have been assigned to the near-side,
and $1<|\Delta\phi|<2\pi-1$ to the away-side, respectively. The 
jet-associated yield $ \frac{1}{N_{T}}\frac{dN_{TA}}{dp_t}$ 
at the near- and away-side, integrated over the respective $\Delta\phi$ ranges
and for sliding windows of the trigger $p_t$ is shown in Fig.~\ref{fig2} 
(left and middle panel). We observe that the spectrum at the near-side is
significantly steeper than at the away-side for $p_t(A) < 1$~GeV/$c$, 
independent of the trigger $p_t$. 
Such behaviour can be explained by the requirement of a hard trigger, which,
biasing the fragmentation, softens
the observed spectrum of associated particles at the 
near-side, whereas the away-side spectrum
remains unbiased. Similar softening is  
observed in calculations using the PYTHIA
event generator~\cite{PYTHIA}, 
corroborating the fragmentation picture at the near-side.
In detail, the measured $p_t$ dependence of 
this effect reveals distinct differences to PYTHIA. 
This is demonstrated in Fig.~\ref{fig2} 
(right panel), where
the ratio of the away- to near-side yield is compared to the same quantity 
from PYTHIA.
For $p_t< 2$~GeV/$c$, the data indicate an enhanced yield at the away-side as 
compared
to the expectation from vacuum fragmentation implied by PYTHIA. 

%%% figure 2.....

\begin{figure}[ht]
\begin{center}
\includegraphics[width=10.cm]{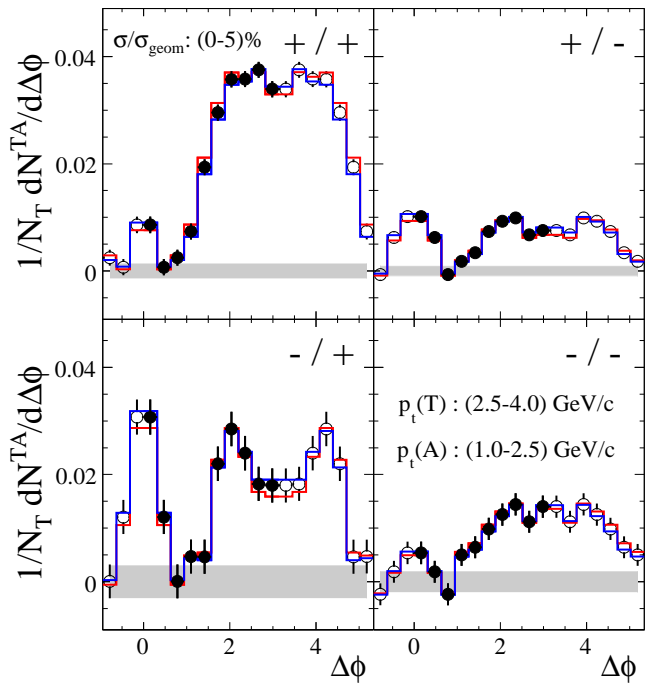}
   \caption{(Color online) 
	Jet-associated yield per trigger for different trigger/associated charge
	combinations in central Pb-Au collisions at 158$A$ GeV/$c$. The open symbols are reflected
	about $\Delta\phi = \pi$. The blue
	and red lines and histograms, and the grey band indicate 
	the systematic uncertainty 
	in the flow subtraction and determination of the $b$-parameter, 
	respectively.}
\label{fig3}
\end{center}
\end{figure}

A more detailed view of the possible mechanisms of jet modifications
can be provided by investigating the charge dependence of two-particle
correlations. The conditional jet-associated yields for different charge 
combinations of trigger
and associated particles in central events are shown in Fig.~\ref{fig3}. 
All charge combinations show
a narrow, jet-like peak at the near-side and a broad structure at the away-side,
with a pronounced dip around $\Delta \phi = \pi$ for the unlike-sign charge
combinations.
Moreover, we observe dramatic differences of the relative yields. At the near-side,
unlike-sign combinations are more abundant than like-sign combinations with the
same trigger charge. This is in agreement with local charge conservation in the
jet fragmentation process. At the away-side, positive associates are more abundant
than negative ones, for both trigger charges. 
This asymmetry is also observed in PYTHIA and characterizes
hard scattering at SPS energies as being dominated by large-$x$ partons, hence 
reflecting the positive net charge of valence quarks. We have verified with 
PYTHIA that this asymmetry vanishes at RHIC energies and beyond, consistent with
observation~\cite{star2,phe7}.

%%% figure 3.....

\begin{figure}[ht]
\begin{center}
\includegraphics[width=10.cm]{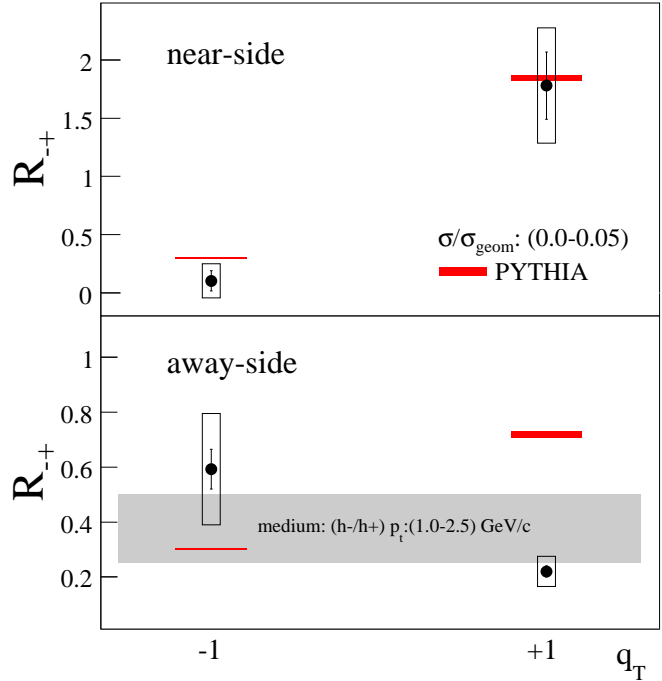}
   \caption{(Color online) Ratio of negative to positive associated
	particles $R_{-+}$ at the near- and away-side in central 
	Pb-Au collisions at 158$A$~GeV/$c$.
	Results as function of the trigger charge $q_T$ are shown in 
 	comparison to PYTHIA calculations. The statistical uncertainties 
	of the PYTHIA calculations are reflected in the thickness of the
	lines.
	Indicated as a shaded band is the inclusive $h^-/h^+$ ratio in the
	associated $p_t$ range.}
\label{fig4}
\end{center}
\end{figure}

As a next step, we study the ratio $R_{-+}$ of the observed yields of 
negative over positive particles at the near- and the away-side in the 
jet-like component. 
We obtain for a given trigger charge 
\begin{equation}
R_{-+}=\int_{\Delta\phi_1}^{\Delta\phi_2}\hat{J}_2^-(\Delta\phi)d\Delta\phi/
	\int_{\Delta\phi_1}^{\Delta\phi_2}\hat{J}_2^+(\Delta\phi)d\Delta\phi,
\end{equation}
where $\hat{J}_2^-$ and $\hat{J}_2^+$ are the jet-associated yields of 
negatively and positively charged associated particles, respectively. 
Integration limits are as above $\Delta\phi_1 = -1$,
$\Delta\phi_2 = 1$ for the near-side and $\Delta\phi_1 = 1$, 
$\Delta\phi_2 = 2\pi -1$ for the away-side.
The results for $2.5 < p_{t}(T) < 4.0$~GeV/$c$ 
and $1.0 < p_{t}(A) < 2.5$~GeV/$c$ in central Pb-Au collisions 
are shown as function of the trigger charge $q_T$ in Fig.~\ref{fig4}.
As a reference for vacuum fragmentation, we calculate these ratios with PYTHIA. 
To this end, calculations of p-p, n-n, p-n, and n-p collisions 
have been properly weighted according to their occurence in Pb-Au 
collisions, assuming binary collision scaling. 
On the near-side,
the PYTHIA calculations match very well the observed ratios $R_{-+}$ for
negative and positive triggers, as seen in the upper panel in Fig.~\ref{fig4}. 
This supports the view that the near-side correlations are characterized
by vacuum fragmentation of jets. 
At the away-side, we compare our results also to the 
inclusive charge ratio of bulk hadrons $(h^-/h^+)$ in the associate 
$p_t$ range,
indicated by the shaded band in Fig.~\ref{fig4}. For positive triggers, the
observed ratio $R_{-+}$ is a factor three smaller than the PYTHIA 
calculation. 
Remarkably, the measurement 
is much closer to the medium charge ratio, 
suggesting that the jet energy has been transfered to the medium
prior to hadronization. Incidentally, the PYTHIA prediction coincides with the
medium charge ratio for negative trigger particles, thus allowing no discrimination
between vacuum fragmentation and energy deposition to the medium. The measurement
for negative trigger particles is in agreement with both scenarios.

%%% figure 4.....

To summarize, we have presented results of a high-$p_t$ 
two-particle correlation analysis in Pb-Au collisions at SPS. 
The data exhibit a narrow peak at the near-side, indicative of jet correlations,
and a broad structure at the away-side. In central collisions,
we observe a significant dip of the two-particle yield at $\Delta \phi = \pi$,
which is most pronounced for unlike-sign charge combinations.
The $p_t$ spectrum of associated particles at the near-side is softer than
at the away-side, in accordance with the expectation of a trigger bias and
in qualitative agreement with PYTHIA. However, a more detailed comparison to PYTHIA 
discloses an excess of soft particles ($p_t < 2$~GeV/$c$) on the away-side
and suggests it is created
in response to the energy deposition of the hard-scattered parton.
The study of the charge dependence of the two-particle yield and a comparison
to PYTHIA reveals that the near-side correlation, in particular its charge
composition, is clearly jet-like. In contrast, the away-side charge composition
is inconsistent with vacuum fragmentation represented by PYTHIA. 
It is, however, in agreement with the bulk charge composition in this $p_t$
range, suggesting that the correlated particles do not stem from fragmentation
directly but emerge due to a local energy deposition of the parton,
implying a medium-modified fragmentation function.
In essence, the present data demonstrate that jet properties in nucleus-nucleus
collisions are modified at SPS energies, indicating that also at these
energies matter of considerable opaqueness is created. The charge composition
and transverse momentum dependence of the away-side yield are consistent
with a scenario including significant energy transfer of the parton to the medium.

\section{Acknowledgements} \nonumber

This work was supported by GSI-F\&E, the German BMBF, the Virtual Institute VI-SIM
and the ExtreMe Matter Institute EMMI of the German Helmholtz Association,
the Israel Science Foundation,
the Minerva Foundation and by the Grant Agency and Ministry of Education
of the Czech Republic.

\end{document}